\documentclass[leqno]{article}
\usepackage{makeidx}

\usepackage{comment}
\usepackage{epsfig} \usepackage{xspace}

\usepackage{amsmath}
\usepackage{amssymb}

\let\ensuremathTEMP\ensuremath

\makeatletter%
\def\nottoobig#1{{\hbox{$\left#1\vcenter
to1.111\ht\strutbox{}\right.\n@space$}}}
\makeatother%

\makeatother%

\makeatletter%[

\newcount\hour \newcount\minutes \hour=\time \divide\hour by 60
\minutes=\hour \multiply\minutes by -60 \advance\minutes by \time
\def\mmmddyyyy{\ifcase\month\or Jan\or Feb\or Mar\or Apr\or May\or
Jun\or Jul\or Aug\or Sep\or Oct\or Nov\or Dec\fi \space\number\day,
\number\year}
\def\hhmm{\ifnum\hour<10 0\fi\number\hour :%
  \ifnum\minutes<10 0\fi\number\minutes} \def\Draft{{\it Draft of
\mmmddyyyy}}

\def\ps@jtsheadings{%
\def\@oddhead{\it\rightmark\hfil\rm\thepage}%
\def\@oddfoot{\hfil\Draft}%
\if@twoside%[
\def\@evenhead{\rm\thepage\hfil\it\leftmark}%
\def\@evenfoot{\Draft\hfil}%
\else
\let\@evenhead\@oddhead%
\let\@evenfoot\@oddfoot%
\fi%]
}
\def\ps@jtsplain{%
\def\@oddhead{\hfil\Draft}%
\def\@oddfoot{\hfil\rm\thepage\hfil}%
\let\@evenfoot\@oddfoot%
\if@twoside \def\@evenhead{\Draft\hfil} \else \let\@evenhead\@oddhead
\fi }

\def\chaptermark#1{\markboth{\thechapter.\ #1}{\thechapter.\ #1}}%
\def\sectionmark#1{\markright{\thesection.\ #1}}

\def\section{\@startsection {section}{1}{\z@} {3.5ex plus1ex
    minus.2ex}{2.3ex plus.2ex}{\Large\bf}}
    \def\subsection{\@startsection{subsection}{2}{\z@} {3.25ex plus1ex
    minus.2ex}{1.5ex plus.2ex}{\large\bf}}
    \def\subsubsection{\@startsection{subsubsection}{3}{\z@} {3.25ex
    plus1ex minus.2ex}{1.5ex plus.2ex}{\normalsize\bf}}
    \def\paragraph{\@startsection{paragraph}{4}{\z@} {3.25ex plus1ex
    minus.2ex}{1em}{\normalsize\bf}}
    \def\subparagraph{\@startsection{subparagraph}{4}{\parindent}
    {3.25ex plus1ex minus.2ex}{1em}{\normalsize\bf}}

\makeatother%]

\makeatletter \@beginparpenalty=10000 \makeatother
\def\underl#1 {\leavevmode\let\first=\relax\underli #1 }
\def\underli#1 {\ifx&#1\let\next=\relax\unskip
\else\let\next=\underli\first\ulinebox{#1}\fi\let\first=\undersp\next}
\def\undersp{\penalty50\ulinebox{\space}\penalty50}
\def\ulinebox#1{\vtop{\hbox{\strut#1}\hrule}}%Taken from TexHax v.88, 52
\def\unice#1 {\underl #1 & }
\def\desclabel#1{\bf #1\hfil}
\def\desc{\list{}{%
\setlength{\leftmargin}{0pt} \labelwidth= \leftmargin \advance
\labelwidth by -\labelsep \let \makelabel=\desclabel}}

\def\descHACKlabel#1{\bf #1\hfil}
\def\descHACK{\list{}{%
\setlength{\leftmargin}{0pt} \labelwidth= \leftmargin \advance
\labelwidth by -\labelsep \let \makelabel=\descHACKlabel}}

\newcounter{extremeleftlistcounter}
  {\begin{list}{\arabic{extremeleftlistcounter}~~~}{\usecounter{extremeleftlistcounter}%
        \setlength{\labelsep}{0pt}\setlength{\leftmargin}{0pt}%
        \setlength{\labelwidth}{0pt}\setlength{\listparindent}{0pt}}}%
  {\end{list}}

\newcounter{leftlistcounter}
  {\begin{list}{\arabic{leftlistcounter}~~~}{\usecounter{leftlistcounter}%
        \setlength{\labelsep}{0pt}\setlength{\leftmargin}{15pt}%
        \setlength{\labelwidth}{15pt}\setlength{\listparindent}{0pt}}}%
  {\end{list}}



\makeatletter %{
\def\union{\,\cup} \def\inter{\,\cap}

\newlength{\filength} 
\newsavebox{\gcbox}
\sbox{\gcbox}{\framebox[\filength]{\rule{0ex}{2ex}}}

\newlength{\leftjustindent} \newlength{\@leftjustindent}
\def\leftjust{\let\\\@leftjustcr\let\end\@endleftjust
\addtolength{\@leftjustindent}{\leftjustindent} \vcenter\bgroup
\halign\bgroup \hbox to\displaywidth{
\rule{\@leftjustindent}{0ex}$\displaystyle##$\hfill }\crcr }
\def\endleftjust{\crcr\egroup\egroup\endgroup}
\def\@endleftjust#1{\crcr\egroup\egroup\@checkend{#1}\endgroup}
\def\@leftjustcr{\crcr}

\newtheorem{theorem}{Theorem}[section]

\newtheorem{corollary}[theorem]{Corollary}
\newtheorem{claim}[theorem]{Claim}

\newtheorem{lemma}[theorem]{Lemma}

\newcommand{\singlespacing}{\let\CS=
\@currsize\renewcommand{\baselinestretch}{1}\tiny\CS}
\newcommand{\singlespacingplus}{\let\CS=
\@currsize\renewcommand{\baselinestretch}{1.25}\tiny\CS}
\newcommand{\doublespacing}{\let\CS=
\@currsize\renewcommand{\baselinestretch}{1.75}\tiny\CS}
\newcommand{\draftspacing}{\let\CS=
\@currsize\renewcommand{\baselinestretch}{2.0}\tiny\CS}
\newcommand{\hugedraftspacing}{\let\CS=
\@currsize\renewcommand{\baselinestretch}{2.4}\tiny\CS}

\makeatother%}

\mathcode`\0="0030      % Make the digits roman always
\mathcode`\1="0031 \mathcode`\2="0032 \mathcode`\3="0033
\mathcode`\4="0034 \mathcode`\5="0035 \mathcode`\6="0036
\mathcode`\7="0037 \mathcode`\8="0038 \mathcode`\9="0039
\newtheorem{definition}[theorem]{Definition}

\makeatletter \clubpenalty=\@highpenalty \widowpenalty=\@highpenalty
\makeatother

\makeatletter
\newcommand{\niceonespacing}{\let\CS=\@currsize\renewcommand{\baselinestretch}{1.1}\tiny\CS}\newcommand{\nicetwospacing}{\let\CS=\@currsize\renewcommand{\baselinestretch}{1.2}\tiny\CS}
\newcommand{\nicethreespacing}{\let\CS=\@currsize\renewcommand{\baselinestretch}{1.3}\tiny\CS}
\newcommand{\singlespacingplusplus}{\let\CS=\@currsize\renewcommand{\baselinestretch}{1.35}\tiny\CS}
\newcommand{\nicefourspacing}{\let\CS=\@currsize\renewcommand{\baselinestretch}{1.4}\tiny\CS}
\newcommand{\nicefivespacing}{\let\CS=\@currsize\renewcommand{\baselinestretch}{1.5}\tiny\CS}
\newcommand{\nicesixspacing}{\let\CS=\@currsize\renewcommand{\baselinestretch}{1.6}\tiny\CS}
\makeatother

\makeatletter \def\@cite#1#2{[#1\if@tempswa , #2\fi]} \makeatother

\makeatletter
\def\@citex[#1]#2{\if@filesw\immediate\write\@auxout{\string\citation{#2}}\fi
\def\@citea{}\@cite{\@for\@citeb:=#2\do
{\@citea\def\@citea{,\linebreak[0]}\@ifundefined {b@\@citeb}{{\bf
?}\@warning
       {Citation `\@citeb' on page \thepage \space undefined}}%
\hbox{\csname b@\@citeb\endcsname}}}{#1}} \makeatother

\makeatletter
\def\ps@thesis{\def\@oddhead{\hfil\rm\thepage\hfil}\def\@oddfoot{}\def\@evenhead{\hfil\rm\thepage\hfil}\def\@evenfoot{}\def\chaptermark##1{}\def\sectionmark##1{}}
\makeatother

\makeatletter
\def\foobarpt{\textfont\z@\tenrm \scriptfont\z@\ninrm
  \scriptscriptfont\z@\sevrm \textfont\@ne\tenmi \scriptfont\@ne\ninmi
  \scriptscriptfont\@ne\sevmi \textfont\tw@\tensy
  \scriptfont\tw@\ninsy \scriptscriptfont\tw@\sevsy
  \textfont\thr@@\tenex \scriptfont\thr@@\tenex
  \scriptscriptfont\thr@@\tenex
  \def\unboldmath{\everymath{}\everydisplay{}\@nomath\unboldmath
  \textfont\@ne\tenmi \textfont\tw@\tensy \textfont\lyfam\tenly
  \@boldfalse}\@boldfalse
  \def\boldmath{\@ifundefined{tenmib}{\global\font\tenmib\@mbi\@magscale1\global
  \font\tensyb\@mbsy \@magscale1\global\font
  \tenlyb\@lasyb\@magscale1\relax\@addfontinfo\@xiipt
  {\def\boldmath{\everymath
  {\mit}\everydisplay{\mit}\@prtct\@nomathbold \textfont\@ne\tenmib
  \textfont\tw@\tensyb
                \textfont\lyfam\tenlyb\@prtct\@boldtrue}}}{}\@xiipt\boldmath}%
\def\prm{\fam\z@\tenrm}%
\def\pit{\fam\itfam\tenit}\textfont\itfam\tenit
   \scriptfont\itfam\ninit \scriptscriptfont\itfam\sevit
   \def\psl{\fam\slfam\tensl}\textfont\slfam\tensl
   \scriptfont\slfam\tensl \scriptscriptfont\slfam\tensl
   \def\pbf{\fam\bffam\tenbf}\textfont\bffam\tenbf
   \scriptfont\bffam\ninbf \scriptscriptfont\bffam\ninbf
   \def\ptt{\fam\ttfam\tentt}\textfont\ttfam\tentt
   \scriptfont\ttfam\nintt \scriptscriptfont\ttfam\nintt
   \def\psf{\fam\sffam\tensf}\textfont\sffam\tensf
   \scriptfont\sffam\tensf \scriptscriptfont\sffam\tensf
\def\psc{\@getfont\psc\scfam\@xiipt{\@mcsc\@magscale1}}%
\def\ly{\fam\lyfam\tenly}\textfont\lyfam\tenly \scriptfont\lyfam\ninly
   \scriptscriptfont\lyfam\sevly \@setstrut \rm}

\makeatother

\newcommand{\naturalnumber}{\ensuremath{{ \mathbb{N} }}}

\newcommand{\integers}{\ensuremath{{ \mathbb{Z} }}}

\newcommand{\p}{{\rm P}} 
 \newcommand{\np}{{\rm NP}}

\newcommand{\sproof}{\noindent{\bf Proof}\quad}

\newcommand{\condition}{\,\ensuremathTEMP{\mbox{\large$|$}}\:}

\newenvironment{block}{\begin{list}{\hbox{}}{\leftmargin 1em
    \itemindent -1em \topsep 0pt \itemsep 0pt \partopsep
    0pt}}{\end{list}}

\makeatletter \def\@listI{\leftmargin\leftmargini \parsep 4.5pt plus
1pt minus 1pt\topsep 6pt plus 2pt minus 2pt \itemsep 2pt plus 2pt
minus 1pt}

\let\@listi\@listI \@listi \makeatother

\makeatletter %{
 \newcommand{\setoffdisplay}{\rule{5.9in}{1pt}}

\makeatother


\newcommand{\uc}[2]{\ensuremath{(#1,#2)\hbox{-}{\rm UC}}}
\newcommand{\dc}[2]{\ensuremath{(#1,#2)\hbox{-}{\rm DC}}}
\newcommand{\disjpath}[1]{\ensuremath{{\rm DISJ\hbox{-}PATH}_{#1}}}

\newcommand{\mymod}[2]{\ensuremath{#1~({\rm mod}~#2)}}

\begin{document}
%\singlespacingplus

\title{Complexity of Cycle Length Modularity Problems in Graphs\footnote{Supported in part by grants NSF-INT-9815095/\protect\linebreak[0]DAAD-315-PPP-g\"u-ab, NSF-CCR-0311021, and a DAAD grant.}}
\author{
\emph{Edith Hemaspaandra}\\
Department of Computer Science\\
Rochester Institute of Technology\\
Rochester, NY 14623, USA\\
{\protect\tt{}eh@}\protect\linebreak[0]\mbox{\protect\tt{}cs.rit.edu}.
\and
  \emph{Holger Spakowski}\thanks{
   Work done while
   visiting the University of Rochester.}\\
  Institut f\"{u}r Informatik\\
  Heinrich-Heine-Universit\"{a}t D\"{u}sseldorf\\
  40225 D\"{u}sseldorf, Germany\\
  {\protect\tt{}spakowsk@}\protect\linebreak[0]\mbox{\protect\tt{}cs.uni-duesseldorf.de}.
\and 
  \emph{Mayur Thakur}\\
  Department of Computer Science\\
  University of Rochester\\
  Rochester, NY 14627, USA\\
  {\protect\tt{}thakur@}\protect\linebreak[0]\mbox{\protect\tt{}cs.rochester.edu}.\\
}

\date{June 25, 2003}

% XXX Holger, change to German affiliation.
% XXX Edith, add NSF grant number in final version.

% QUIET LATEX ABOUT SMALL underfill and overfill-ed boxes
%%\typeout{WARNING:  FUZZ used to supress reporting.  Beware.}
% Doesn't WORK!
%%\hfuzz=0.25in
%%\vfuzz=0.25in
\typeout{WARNING: BADNESS used to supress reporting.  Beware.}
\hbadness=3000% The badness above which bad hboxes will be shown
\vbadness=10000 % The badness above which bad vboxes will be shown

%$|\condition |||| \condition \condition \condition$ hello there again
%$$|\condition |||| \condition \condition \condition$$
%\bibliographystyle{/usr/lib/tex/macros/alpha}
\bibliographystyle{alpha}
%\bibliographystyle{elsart-num}
% cover page for tech report
%\pagestyle{empty}

%title page 
%\pagestyle{empty}
%\setcounter{footnote}{0}
%title page spacing
%{\singlespacing\maketitle}
%\vspace{-0.1in}
%abstract spacing
%\niceonespacing

%{\abstract
%}

\maketitle
{\abstract
The even cycle problem for both undirected~\cite{tho:j:disjoint-subgraphs} and
directed~\cite{rob-sey-tho:j:pfaffian} graphs has been the topic of intense
research in the last decade.  In this paper, we study the computational
complexity of \emph{cycle length modularity problems}.  Roughly speaking, in a
cycle length modularity problem, given an input (undirected or directed) graph,
one has to determine whether the graph has a cycle $C$ of a specific length (or
one of several different lengths), modulo a fixed integer.  We denote the two
families (one for undirected graphs and one for directed graphs) of problems by
$\uc{S}{m}$ and $\dc{S}{m}$, where $m \in \naturalnumber$ and $S \subseteq \{0,
1, \ldots, m-1\}$.  $\uc{S}{m}$ (respectively, $\dc{S}{m}$) is defined as
follows: Given an undirected (respectively, directed) graph $G$, is there a
cycle in $G$ whose length, modulo $m$, is a member of $S$?  In this paper, we
fully classify (i.e., as either polynomial-time solvable or as $\np$-complete)
each problem $\uc{S}{m}$ such that $0 \in S$ and each problem $\dc{S}{m}$ such
that $0 \notin S$.  We also give a sufficient condition on $S$ and $m$ for the
following problem to be polynomial-time computable: $\uc{S}{m}$ such that $0
\notin S$. 
}

%\normalspacing
%\draftspacing

%%%%%%%%%%%%%%%%%%%%%%%%%%%%% START:  ABSTR.TEX
%%%%%%%%%%%%%%%%%%%%%%%%%%%%% END:  ABSTR.TEX

%\clearpage
% BODY OF PAPER'S SPACING
%\normalspacing
%\singlespacingplus
%\nicefivespacing
%\draftspacing
%end of page zero
%\pagestyle{empty}
%\pagestyle{plain}

\sloppy

% just for books %  \include{chap-preface}
% just for books %  
% just for books %  
% just for books %  % Table of contents
% just for books %  \cleardoublepage
% just for books %  \typeout{Table of Contents}
% just for books %  \label{contents}
% just for books %  \singlespacing
% just for books %  \tableofcontents\addcontentsline{toc}{chapter}{Contents}
% just for books %  \normalspacing
% just for books %  
% just for books %  % List of figures
% just for books %  \clearpage
% just for books %  \typeout{List of Figures}
% just for books %  \label{lof}
% just for books %  \singlespacing
% just for books %  \listoffigures\addcontentsline{toc}{chapter}{List of Figures}
% just for books %  
% just for books %  % List of tables
% just for books %  \clearpage
% just for books %  \typeout{List of Tables}
% just for books %  \label{lot}
% just for books %  \singlespacing
% just for books %  \listoftables\addcontentsline{toc}{chapter}{List of Tables}
% just for books %  \normalspacing
% just for books %  
% just for books %  
% just for books %  \normalspacing
% just for books %  
% just for books %  \clearpage
% just for books %  
% just for books %  
% just for books %  
% just for books %  
% just for books %  \setcouter{page}{1}
% just for books %          
% just for books %  
% just for books %  
% just for books %  %\pagestyle{cuthdraft}
% just for books %  \pagestyle{headings}
% just for books %  %\pagestyle{thesis}
% just for books %  %\pagestyle{cuthheaders}
% just for books %  \pagenumbering{arabic}
% just for books %  

\section{Introduction}
\noindent
In this paper we study the complexity of problems related to lengths of cycles,
modulo a fixed integer, in undirected and directed graphs.  Given $m \in
\naturalnumber$, and $S \subseteq \{0,1,\ldots,m-1\}$, we define the following
two cycle length modularity problems.

\noindent
$\uc{S}{m} = \{G \condition G$ is an undirected graph such that there exists an
$\ell \in \naturalnumber$ such that $\ell \bmod m \in S$, and there exists a
cycle of length $\ell$ in $G\}$.\\ $\dc{S}{m} = \{G \condition G$ is an directed
graph such that there exists an $\ell \in \naturalnumber$ such that $\ell \bmod
m \in S$, and there exists a directed cycle of length $\ell$ in $G$\}.

The most basic cases of cycle length modularity problems are the following
problems for undirected (respectively, directed) graphs: deciding whether a
given undirected (respectively, directed) graph has a cycle of odd length, and
deciding whether a given undirected (respectively, directed) graph has a cycle
of even length.  We will refer to these problems as the odd cycle problem for
undirected (respectively, directed) graphs, and the even cycle problem for
undirected (respectively, directed) graphs, respectively.  In our notation,
these problems are denoted by $\uc{\{1\}}{2}$ (respectively, $\dc{\{1\}}{2}$)
and $\uc{\{0\}}{2}$ (respectively, $\dc{\{0\}}{2}$).  All these four problems
are now known to be in $\p$.
  
The odd cycle problem and the even cycle problem are quite different in nature.
The reason is that if a closed walk of odd length is decomposed into cycles,
then there is at least one odd cycle in the decomposition.  The corresponding
statement for even walks and even cycles is not true.  Since odd closed walks
can easily be found in polynomial time, it is easy to detect odd cycles.

It is well known that an undirected graph has an odd cycle if and only if the
graph is not bipartite.
% The odd cycle problem for undirected (respectively, directed) graphs is well known to be equivalent to the problem of deciding whether a given undirected (respectively, directed) graph is non-bipartite.  
No such simple characterization is known for the case of even cycles.  However,
Thomassen~\cite{tho:j:disjoint-subgraphs} showed that the family of cycles of
length divisible by $m$ has the Erd\"{o}s-P\'{o}sa
property~\cite{erd-pos:j:independent-circuits}, and then used results from
Robertson and Seymour~\cite{rob-sey:j:graph-minors-2} to prove that the even
cycle problem for undirected graphs is in $\p$.  In fact, Thomassen proved that,
for each $m \in \naturalnumber$, $\uc{\{0\}}{m}$ is in $\p$.  Even though
Thomassen's graph minor and tree-width approach to solving the even cycle
problem is elegant, it means that the algorithm has the drawback of having huge
constants in its running time.  Arkin, Papadimitriou, and
Yannakakis~\cite{ark-pap-yan:j:modularity} used a simpler approach to give an
efficient algorithm for the even cycle problem for undirected graphs.  Their
algorithm is based on their characterization of undirected graphs that do not
contain even cycles with certain efficiently checkable properties of the
biconnected components of these graphs.  However, unlike Thomassen's approach,
their approach does not seem to generalize beyond the $\uc{\{0\}}{2}$ case to,
say $\uc{\{0\}}{3}$.  A related result from Yuster and
Zwick~\cite{yus-zwi:j:even-cycles} shows that, for each $k$, the problem of
deciding if a given undirectd graph has a cycle of length $2k$ is in $\p$.

It is interesting to note that even though the algorithms for the two odd cycle
problems (i.e., undirected and directed) are similar, neither of the two
algorithms mentioned above (namely, Thomassen~\cite{tho:j:disjoint-subgraphs}
and Arkin, Papadimitriou, and Yannakakis~\cite{ark-pap-yan:j:modularity}) seems
to be able to handle the even cycles case for directed graphs.  However,
Robertson, Seymour, and Thomas~\cite{rob-sey-tho:j:pfaffian} (the conference
version of the paper is by McCuaig et al.~\cite{mcc-rob-sey-tho:c:pfaffian})
prove, via giving a polynomial time algorithm for the problem of deciding
whether a bipartite graph has a Pfaffian orientation, that the even cycle
problem for directed graphs is also in $\p$.  We note that Vazirani and
Yannakakis~\cite{vaz-yan:j:pfaffian} proved the polynomial-time equivalence of
the even cycle problem for directed graphs with the following problems:
\begin{enumerate}
\item The problem of checking whether a bipartite graph has a Pfaffian
orientation~\cite{kas:b:graph-theory-crystal-physics},
\item Polya's problem~\cite{pol:j:aufgabe}, i.e., given a square $(0,1)$ matrix
$A$, is there a $(-1,0,1)$ matrix $B$ that can be obtained from $A$ by changing
some of the $1$'s to $-1$'s such that the determinant of $B$ is equal to the
permanent of $A$, and
\item Given a square matrix $A$ of nonnegative integers, determine
      if the determinant of $A$ equals the permanent of $A$. 
\end{enumerate}

Until the even cycle problem was shown to be in $\p$ by
Thomassen~\cite{tho:j:disjoint-subgraphs} (for undirected graphs) and Robertson,
Seymour, and Thomas~\cite{rob-sey-tho:j:pfaffian} (for directed graphs) and even
since then, a lot of interesting research on the even cycle problem and related
problems led to the study of other cycle length modularity problems.  Some
problems have been shown to be in $\p$, while others have been shown to be
$\np$-complete.  We mention some that have a close relationship with the
problems studied in this paper.  As stated earlier,
Thomassen~\cite{tho:j:disjoint-subgraphs} proved that, for each $m \in
\naturalnumber$, the problem of deciding whether an undirected graph has a cycle
of length $\equiv \mymod{0}{m}$ is in $\p$.  That is, for each $m \in
\naturalnumber$, $\uc{\{0\}}{m} \in \p$.  Arkin, Papadimitriou, and
Yannakakis~\cite{ark-pap-yan:j:modularity} study, among other problems, the
problem of deciding whether a directed graph contains cycles of length
$\mymod{p}{m}$.  They prove, via reduction from the directed subgraph
homeomorphism problem~\cite{for-hop-wyl:j:subgraph-homeomorphism}, that for all
$m > 2$ and for all $p$ such that $0 < p < m$, $\dc{\{p\}}{m}$ is
$\np$-complete.  
They also give a polynomial-time algorithm for
the problem of finding the greatest common
divisor of all cycles in graphs, a problem motivated by the problem of finding
the period of a Markov chain.  Furthermore, they prove that, for all $m > 2$ and for all $0 <
p < m$, the problem of deciding whether all cycles in an undirected graph are of
length $\mymod{p}{m}$ is in $\p$.  %Thus, they prove that the problem of finding
%the greatest common divisor of all cycles in a graph is solvable in polynomial
%time.  
Galluccio and Loebl~\cite{gal-loe:j:prescribed} study the complexity of
the corresponding problem in directed graphs.  They prove that for the case of
planar directed graphs, checking whether all cycles in the input graph are of
length $\mymod{p}{m}$ can be done in polynomial time.

In this paper, we resolve the complexity (as either in $\p$ or $\np$-complete)
of the following problems:
\begin{enumerate}
\item\label{problem:dcsm-0-out} $\dc{S}{m}$, where $0 \not\in S$, and
\item\label{problem:ucsm-0-in} $\uc{S}{m}$, where $0 \in S$.
%\item\label{problem:dc-all-except-one} $\uc{\{0,\ldots,m\} - \{r\}}{m}$, where $0 \leq r < m$.
\end{enumerate}

%We prove that each problem in~\ref{problem:ucsm-0-in} and \ref{problem:dc-all-except-one} is in $\p$.  
We prove that each problem in~\ref{problem:ucsm-0-in} is in $\p$.
For~\ref{problem:dcsm-0-out}, we classify each problem as either in $\p$ or
$\np$-complete, depending only on the properties of $S$ and $m$: If there exist
$0 \le d_1, d_2 < m$ such that $d_1, d_2 \not\in S$ and $(d_1+d_2) \bmod m \in
S$, then $\dc{S}{m}$ is $\np$-complete, otherwise $\dc{S}{m}$ is in $\p$.
We also prove a sufficient condition for $\uc{S}{m}$ (with $0 \notin S$) to be
polynomial-time computable: for each $p \in S$, and for each $d_1, d_2$ such
that $0 \leq d_1, d_2 < m$ and $d_1 + d_2 \equiv \mymod{p}{m}$, it holds that
either $d_1 \in S$ or $d_2 \in S$.  Note that this condition is exactly the same
as that for the ``in $\p$'' case of~\ref{problem:dcsm-0-out} given above.

%%% here goes the organization of the paper
The paper is organized as follows.  In Section~\ref{sec:def}, we introduce the
definitions and notations that will be used in the rest of the paper.  In
Section~\ref{sec:directed}, we present results for cycle length modularity
problems in directed graphs and in Section~\ref{sec:undirected} we present
results for cycle length modularity problems in undirected graphs.  Finally, in
Section~\ref{sec:conclusion} we present some open problems and future research
directions.

\section{Definitions and Notations}\label{sec:def}

In this section we describe the notations used in the rest of the paper.  For
each finite set $S$, let $||S||$ denote the cardinality of $S$.

Since this paper is about graphs, let us make the notations very clear.  An
\emph{undirected graph} $G$ is a pair $(V,E)$, where $V$ is a finite set (known
as the set of vertices or nodes) and $E \subseteq V \times V$ (known as the set
of edges) with the following properties.
\begin{enumerate}
\item For each $u, v \in V$, if $(u,v) \in E$, then $(v,u) \in E$.
\item For each $v \in V$, $(v,v) \not\in E$, that is, self-loops are not
allowed.
\end{enumerate}
A \emph{directed graph} $G$ is a pair $(V,E)$, where $V$ is a finite set and $E
\subseteq V \times V$.  For each graph $G$, let $V(G)$ denote the set of
vertices of $G$, and let $E(G)$ denote the set of edges of $G$.
\begin{comment}
For each graph $G$, and each $1 \leq i \leq ||V_G||$, let $v_i^G$ denote the
$i$-th node in $G$.  The \emph{adjacency matrix} of an undirected (respectively,
directed) graph $G$, denoted $A_G$, is the square binary (i.e., each entry is
either 0 or 1) matrix, of size $||V_G||$ by $||V_G||$, such that, for each $1
\leq i,j \leq ||V(G)||$, $A_G(i,j)$, the entry in the $i$-th row and $j$-th
column, is $1$ if and only if $(v_i^G, v_j^G) \in E$.
\end{comment}
A \emph{walk} of length $k$ in a graph $G$ is a sequence of vertices $(u_0, u_1,
\ldots, u_k)$ with $k \geq 1$ in $G$ such that, for each $0 \leq i < k$, $(u_i,
u_{i+1}) \in E(G)$.
%%%For each path $P = (u_0, u_1, \ldots, u_k)$, $k$ is known as the length of path $P$, denoted $\length(P)$.
% A \emph{sub-path} $P'$ of a path $P = (u_0, u_1, \ldots, u_k)$ in an undirected (respectively, directed) is a subsequence of $P$.  That is, $P'$ is a subpath of $P = (u_0, u_1, \ldots, u_k)$ if and only if there exist $i,j \in \naturalnumber$ with $0 \leq i < j \leq k$ such that $P'$ is the sequence $(u_i, u_{i+1}, \ldots, u_j)$. 
A \emph{path} is a walk where all vertices are distinct.
A \emph{closed walk} in a graph $G$ is a walk $(u_0, u_1, \ldots, u_k)$ in $G$
such that $u_0 = u_k$.  A cycle in an undirected graph is a closed walk $(u_0,
u_1, \ldots, u_{k-1},u_0)$ of length $\geq 3$ such that $u_0, u_1, \ldots,
u_{k-1}$ are $k$ distinct vertices.  A cycle in a directed graph is a closed
walk $(u_0, u_1, \ldots, u_{k-1},u_0)$ such that $u_0, u_1, \ldots, u_{k-1}$ are
$k$ distinct vertices.
% A \emph{cycle} $C$ in a directed graph $G$ is a closed walk $(u_0, u_1, \ldots, u_k)$ in $G$ such that no sub-path of $C$ of length less than $k$ is a closed walk.
% A path $P$ is a \emph{simple path} (equivalently, $P$ is \emph{acyclic} or cycle-free) if no sub-path of $P$ is a cycle.
It should be noted that this definition of a cycle is sometimes called a {\em
simple cycle}.

%%note (mayur): commenting this part according to referee comments
\begin{comment}
Next we define notation related to the $\bmod$ operator.  The $\bmod$ operator
is a binary function defined as follows.  For each $m \in \naturalnumber$ such
that $m > 0$, and for each $i \in \naturalnumber$, $i \bmod m$ is a $j \in
\naturalnumber$ such that $0 \leq j < m$ and such that there is a $b \in
\naturalnumber$ such that $i = b * m + j$.  For each $m$, the $\bmod$ operator
defines an equivalence relation defined over $\naturalnumber$ as follows.  For
each $i, j \in \naturalnumber$, $i \equiv \mymod{j}{m}$ if and only if $i \bmod
m = j \bmod m$.

$\p$ is the class of all languages accepted by some deterministic
polynomial-time Turing machine.  $\np$ is the class of all languages accepted by
some nondeterministic polynomial-time Turing machine.

%%% note (mayur): commenting this part...i think this definition is best stated
%%% when we are actually using it...we only use it once
We denote by $M(n) = O(n^{2.376})$ the complexity of boolean matrix multiplication.
\end{comment}

\section{Cycle Length Modularity Problems in Directed Graphs}\label{sec:directed}
\begin{comment}
\begin{definition}
   Let $q\ge 1$ be a natural number, and $S\subseteq \{ 0,\ldots , q-1\}$. We
   define the following decision problems:
   \begin{itemize}
     \item $\uc{S}{m} $:\\ Input: An undirected graph $G$.\\ Question: Does $G$
          have a (simple) cycle of length $\ell$ such that $(\ell \bmod q) \in S$?
     \item $\dc{S}{m} $:\\ Input: A directed graph $G$.\\ Question: Does $G$
          have a directed (simple) cycle of length $\ell$ such that $(\ell \bmod q)
          \in S$?
   \end{itemize}
\end{definition}
\end{comment}

In this section, we study the complexity of cycle length modularity problems in
directed graphs.  Arkin, Papadimitriou, and
Yannakakis~\cite{ark-pap-yan:j:modularity} proved that, for each $m \in
\naturalnumber$ and each $0 < r < m$, $\dc{\{r\}}{m}$ is $\np$-complete.  In
Theorem~\ref{thm:dir-0-out}, we generalize their result.  For each $m$ and $S$
such that $0 \notin S$, we give a condition on $S$ and $m$ for
$\dc{S}{m}$ to be $\np$-complete.  Furthermore, we prove that if the stated
conditions on $S$ and $m$ are not satisfied, then $\dc{S}{m}$ is in $\p$.

\begin{theorem} \label{thm:dir-0-out}
%%% note(mayur): changing ``S \subseteq {0,..,m-1} with 0 \notin S'' to ``S
%%% \subseteq {1,..,m-1}''  according to referee comments
   For all $m\ge 1$ and $S\subseteq \{1, \ldots , m-1\}$, the
   following is true:
   \begin{description}
      \item [(i)] If there is a $p\in S$, and $d_1 \notin S, d_2 \notin S$ such
            that $0\le d_1, d_2 < m$ and $d_1 + d_2 \equiv p \pmod{m}$, then
            $\dc{S}{m}$ is $\np$-complete.
      \item [(ii)] Otherwise, $\dc{S}{m}$ is in $\p$.
   \end{description}
\end{theorem}

\sproof
To prove (i), let $p \in S$ and $d_1, d_2 \notin S$ be such that $0\le d_1, d_2
< m$ and $d_1 + d_2 \equiv \mymod{p}{m}$.  We closely follow the proof of
Theorem 1 in Arkin, Papadimitriou, and
Yannakakis~\cite{ark-pap-yan:j:modularity}.  Fortune, Hopcroft, and
Wyllie~\cite{for-hop-wyl:j:subgraph-homeomorphism} showed that the directed
subgraph homeomorphism problem is NP-complete for any fixed directed graph that
is not a tree of depth 1.  In particular, the following problem is NP-complete:
\begin{quote}
Given a directed graph $G$ and vertices $s$ and $t$ in $V(G)$, does $G$ contain
a cycle through both $s$ and $t$?
\end{quote}
We will prove that $\dc{S}{m}$ is $\np$-complete by polynomial-time many-one
reducing this problem to $\dc{S}{m}$.  We now specify a polynomial-time function
$\sigma$ that reduces this
%the directed subgraph homeomorphism 
problem to $\dc{S}{m}$.  Given a directed graph $G$ and vertices $s, t \in
V(G)$, $\sigma(\langle G, s, t\rangle)$ outputs the graph $G'$ where $G' =
(V',E')$ is defined as follows.
%%$V' = \{v_i^0 \condition 1 \leq i \leq n\} \union \{v_{i,j}^k \condition 1 \leq i, j \leq n \land 1 \leq k \leq m-1\}$.
$V'$ and $E'$ are constructed in the following steps.  (Note that in the steps
below, we can assume that $d_1 \neq 0$ and $d_2 \neq 0$, because if either $d_1$
or $d_2$ is equal to $0$, then the preconditions of (i) cannot be satisfied.)
\begin{enumerate}
\item Set $V' := V$.  Set $E' := \emptyset$.
\item For every edge $(v,s) \in E(G)$, do the following.
\begin{enumerate}
\item Set $V' := V' \union \{w_j \condition 1 \leq j \leq d_1-1\}$, where the
$w_j$'s are new vertices.
\item Set $E' := E' \union \{(v, w_1), (w_1, w_2), \ldots,
(w_{d_1-1}, s)\}$.
\end{enumerate}
\item For every edge $(v,t) \in E(G)$, do the following.
\begin{enumerate}
\item Set $V' := V' \union \{w_j \condition 1 \leq j \leq d_2-1\}$, where the
$w_j$'s are new vertices.
\item Set $E' := E' \union \{(v, w_1), (w_1, w_2), \ldots,
(w_{d_2-1}, t)\}$.
\end{enumerate}
\item For every edge $(v,w) \in E(G)$ such that $v, w \not \in \{s,t\}$, do the
following.
\begin{enumerate}
\item Set $V' := V' \union \{w_j \condition 1 \leq j \leq m-1\}$, where the
$w_j$'s are new vertices.
\item Set $E' := E' \union \{(v, w_1), (w_1, w_2), \ldots, 
(w_{m-1}, w)\}$.
\end{enumerate}
\end{enumerate}
It is easy to see that the cycles in $G'$ have the following properties.
\begin{enumerate}
\item All cycles in $G'$ going through neither $s$ nor $t$ have length $\equiv
\mymod{0}{m}$.
\item All cycles in $G'$ going through $s$ but not through $t$ have length
$\equiv \mymod{d_1}{m}$.
\item All cycles in $G'$ going through $t$ but not through $s$ have length
$\equiv \mymod{d_2}{m}$.
\item All cycles in $G'$ going through $s$ and $t$ have length $\equiv
\mymod{(d_1 + d_2)}{m} \equiv \mymod{p}{m}$
\end{enumerate}

Roughly speaking, we replace each edge $e \in E(G)$ that ends in $s$, by a
series of $d_1$ edges in $G'$ such that the series of edges ends in $s$.
Similarly, we replace each edge $e \in E(G)$ that ends in $t$, by a series of
$d_2$ edges in $G'$ that ends in $t$.  It is clear from the construction of $G'$
that there is a cycle through $s$ and $t$ in $G$ if and only if there is a cycle
through $s$ and $t$ in $G'$.  Since $\{0, d_1, d_2\} \inter S = \emptyset$ and
$p \in S$, it follows from the properties stated above that there is a cycle
through $s$ and $t$ in $G$ if and only if there is a cycle of length $\equiv
\mymod{p}{m}$ in $G'$.  Also, it is clear that $G'$ can be computed from $G$ in
polynomial time.  It follows that $\dc{S}{m}$ is $\np$-hard.  Note that, for
each $S$ and $m$, $\dc{S}{m}$ is clearly in $\np$.  Thus, $\dc{S}{m}$ is
$\np$-complete.

We will now prove (ii).  Let $m \geq 1$ and $S\subseteq \{1, \ldots , m-1\}$ be such
that for all $p\in S$ and all $d_1, d_2$, if $0 \le d_1, d_2 < m$ and $d_1 + d_2
\equiv \mymod{p}{m}$, then $d_1 \in S$ or $d_2\in S$.

We claim that the following algorithm solves $\dc{S}{m}$ in polynomial time:

\medskip

\noindent Input: A directed graph $G$.\vspace{-0.2cm}
\begin{enumerate}
\item {\bf for} each $p\in S$ {\bf do} \vspace{-0.2cm}
\item \hspace*{0.5cm} {\bf if} $G$ has a closed walk of length $\equiv
  p\pmod{m}$ {\bf then} accept. \vspace{-0.2cm}
\item reject.
\end{enumerate}
Clearly, step 2 can be done in polynomial time. If the algorithm rejects, then
obviously $G$ is not in $\dc{S}{m}$.  To complete the proof of (ii), we will
prove the following claim.
\begin{claim}\label{clm:helper}
  If $G$ has a closed walk $W$ of length $\equiv \mymod{p}{m}$ for some $p\in
  S$, then $G$ has a cycle of length $\equiv \mymod{p'}{m}$ for some $p' \in S$.
\end{claim}
\sproof
The proof is by induction on the length of $W$.  The claim is certainly true for
all closed walks $W$ of length $1$.  Assume that the claim is true for all
closed walks $W$ whose length is less than $k$.  Suppose $G$ has a closed walk
$W$ of length $k$ with $k \bmod m = p$ and $p \in S$.  Distinguish the following
two cases.
\begin{description}
\item [Case 1:] $W$ is a cycle.\\ Then we are done.
\item [Case 2:] $W$ is not a cycle.\\ Then there exist $\ell_1 > 0$, $\ell_2 >
0$, $d_1 < m$, and $d_2 < m$ such that $W$ can be decomposed into a simple cycle
$C$ of length $\ell_1$ and a closed walk $W'$ of length $\ell_2$ such that
$\ell_1 \equiv \mymod{d_1}{m}$, $\ell_2 \equiv \mymod{d_2}{m}$.  Since $\ell_1 +
\ell_2 = k$, it follows that $\ell_1 + \ell_2 \equiv d_1 + d_2 \equiv
\mymod{p}{m}$.  We know that $d_1\in S$ or $d_2\in S$.  If $d_1 \in S$ then we
are done.  If $d_2\in S$ we are done by the induction hypothesis.
\end{description}
\noindent Thus, Claim~\ref{clm:helper} holds, and so Theorem~\ref{thm:dir-0-out} holds.
%\eproofof{Claim~\ref{clm:helper}}\\
%\eproofof{Theorem~\ref{thm:dir-0-out}}\\

\medskip
\noindent
As an immediate corollary, we get that the problem of deciding whether all
cycles in a directed graph have length $\equiv \mymod{0}{m}$ is in $\p$.

\begin{corollary}\label{cor:dc-all-except-0}
For each $m \in \naturalnumber$, $\dc{\{1,2,\ldots,m-1\}}{m} \in \p$.
\end{corollary}
We note that Corollary~\ref{cor:dc-all-except-0} also follows from the fact that
finding the period (greatest common divisor of all cycle lengths) of a graph is
in $\p$~\cite{bal-vei:j:period} (see
also~\cite{knu:tr:strong-components,tar:j:dfs}).  %Yuster and
%Zwick~\cite{yus-zwi:j:even-cycles} proved that, for directed graphs, a shortest
%odd length cycle can be found in polynomial time.  Note that the odd cycle
%problem for directed graphs is $\dc{\{1,2,\ldots,m-1\}}{m}$, when $m = 2$.
%Thus, it is interesting to ask whether Yuster and Zwick's result can be
%generalized, i.e., can we find in polynomial time a shortest cycle of length
%$\not\equiv \mymod{0}{m}$.  We note that in fact this is the case: Using an
%algorithm similar to Yuster and Zwick's, we can find a shortest cycle of length
%$\not\equiv \mymod{0}{m}$.

Yuster and
Zwick~\cite{yus-zwi:j:even-cycles} proved that for directed graphs, a shortest
odd length cycle can be found in time $O(||V||\cdot ||E||)$. We show that for all $\dc{S}{m}$-problems
satisfying the condition (ii) of Theorem~\ref{thm:dir-0-out}, a shortest cycle with length,
modulo $m$, in $S$ can be found in time $O(M(|| V||) \cdot\log ||V|| )$, where
$M(n) = n^{2.376}$ is the complexity of boolean matrix multiplication. For the special 
case $m=2$, $S = \{ 1\}$, the algorithm is for dense graphs an improvement over 
the one given in~\cite{yus-zwi:j:even-cycles}.
\begin{theorem}\label{thm:shortest-cycle}
  For all $m\ge 2$ and $S\subseteq \{ 0, \ldots , m-1\}$ with $0\notin S$ the
  following is true: If for all $p\in S$, and all $d_1, d_2$, such that $0\le d_1,
  d_2 < m$ and $d_1 + d_2\equiv \mymod{p}{m}$, it holds that $d_1\in S$ or $d_2\in
  S$, then there is an  $O(M(|| V||) \cdot\log ||V|| )$ time algorithm that computes 
  a shortest cycle $C$ such that
  the length of $C$, modulo m, is in $S$.
\end{theorem}
\sproof
  If the precondition of Theorem~\ref{thm:shortest-cycle} holds, every closed
  walk whose length, modulo $m$, belongs to $S$, is a cycle or 
  decomposes into cycles such that the length of at least one of these cycles, 
  modulo $m$, belongs to $S$. Hence the problem reduces to finding a shortest
  closed walk whose length, modulo $m$, belongs to $S$.

  Let $G=(V,E)$, where $V = \{ v_1, \ldots , v_n\}$. For every 
  $r\in\{ 0, \ldots , m-1\}$ and $0 < k \le n$, we define the boolean matrix $A_{k,r}$
  by $A_{k,r}(i, j) \stackrel{df}{=} 1$ iff there is a walk of length $\ell$ 
  from $v_i$ to $v_j$ in $G$ with
  $0<\ell\le k$ and $\ell\equiv r\pmod{m}$.
  With $O(\log n)$ boolean matrix multiplications we can
  determine $k_{min}$, the length of the desired closed walk.
  The value of $k_{min}$ equals the smallest $k$ with
  $A_{k, r'}(i,i) = 1$ for some $i\in\{ 1,\ldots , n\}$ and $r'\in S$.
  First, compute the matrices $A_{k,r}$ where $k$ is a power of $2$, using the 
  identity
  \begin{displaymath}
    A_{2k, r} = \bigvee_{i=0}^{m-1}(A_{k, i} \wedge A_{k,r-i}) \vee A_{k, r},
  \end{displaymath}
  where $\wedge$ and $\vee$ stand for boolean matrix multiplication and componentwise
  'or', respectively.
  Note that $A_{1,1}$ is the adjacency matrix of $G$, and $A_{1,r}$, $r\not= 1$,
  is a zero matrix. After that, apply binary search to determine $k_{min}$,
  and a representation of $A_{k_{min}, r'}$ as product of matrices $A_{k,r}$ with
  $k$ being a power of $2$. A specific closed walk  
  with length $k_{min}$ (which we know, is a cycle)   can now easily be found
  in additional $O(|| V|| ^2)$ time.

\section{Cycle Length Modularity Problems in Undirected Graphs}\label{sec:undirected}
In this section, we study the complexity of problems $\uc{S}{m}$, for different
$S$ and $m$.  The case when $S = \{0\}$ has been shown to be in $\p$ by
Thomassen~\cite{tho:j:disjoint-subgraphs}.  We extend Thomassen's result and
prove that for all $S$ such that $0 \in S$, $\uc{S}{m}$ is in $\p$.

\begin{theorem}\label{thm:ucsm-0-in}
For each $m$, and each $S \subseteq \{0, \ldots, m-1\}$ such that $0 \in S$,
$\uc{S}{m} \in \p$.
\end{theorem} 

The proof of Theorem~\ref{thm:ucsm-0-in} is an extension of the proof of
Thomassen's result for $\uc{\{0\}}{m}$, which in turn is based on the result
from Robertson and Seymour~\cite{rob-sey:j:graph-minors-2} for the $k$-disjoint
paths problem.  We will need the following results related to tree-widths for
the proof of Theorem~\ref{thm:ucsm-0-in}.  Tree-width is an invariant of graphs
that has been a central concept in the development of algorithms for fundamental
problems in graph theory.  See~\cite{rob-sey:j:graph-minors-2} for a definition
of tree-width, and~\cite{rob-sey:b:survey} for a survey on graph minor results.
We will not define tree-widths because the definition is rather involved and for
the proof of Theorem~\ref{thm:ucsm-0-in} we need know only the following fact
about tree-widths of graphs.

\begin{theorem}$\cite{rob-sey:j:graph-minors-2}$\label{thm:tree-width}
For each $t \in \naturalnumber$, there is a polynomial-time algorithm for
deciding whether an undirected graph has tree-width at least $t$.
\end{theorem}

% In other words, the function mapping an undirected graph to its tree-width is polynomial-time computable. 
The following theorem shows that, for fixed $m$, all graphs of sufficiently
large tree-width have a cycle whose length is a multiple of $m$.

\begin{theorem}$\cite{tho:j:disjoint-subgraphs}$\label{thm:large-tree-width}
For each $m$, there exists a $t_m \in \naturalnumber$ such that, for each
undirected graph $G$ with tree-width at least $t_m$, $G$ contains a cycle of
length $\equiv \mymod{0}{m}$.
\end{theorem}

Roughly speaking, Theorem~\ref{thm:large-tree-width} allows us to handle those
graphs that have large tree-widths.  Theorem~\ref{thm:k-disj-paths} allows us to
handle small tree-widths.

\begin{definition}
For each $t, m \in \naturalnumber$, $d_1, d_2, \ldots d_k$ such that, for each
$1 \leq i \leq k$, $d_i < m$, $\disjpath{\langle t, m, d_1, d_2, \ldots,
d_k\rangle}$ is defined as follows: $\disjpath{\langle t, m, d_1, d_2, \ldots,
d_k\rangle} = \{\langle G, x_1, y_1, \ldots, x_k, y_k \condition G$ is an
undirected graph such that (a) $G$ has tree width at most $t$, (b) for each $i$,
$x_i$ and $y_i$ are vertices in $G$, and (c) there exist $k$ node-disjoint paths
$P_1, P_2, \ldots, P_k$ in $G$ such that, for each $1 \leq i \leq k$, $P_i$ is a
path connecting $x_i$ and $y_i$ such that $P_i$ has length $\mymod{d_i}{m}\}$.
\end{definition}

\begin{theorem}$\cite{tho:j:disjoint-subgraphs}$\label{thm:k-disj-paths}
Let $t, m, d_1, d_2, \ldots, d_k \in \naturalnumber$ be such that, for each $1
\leq i \leq k$, $d_i < m$.  Then, $\disjpath{\langle t, m, d_1, d_2, \ldots, d_k
\rangle}$ is in $\p$.
\end{theorem}

\noindent
{\bf Proof of Theorem~\ref{thm:ucsm-0-in}.}  Let $m$ and $S$ be such that $m
\in \naturalnumber$, $S \subseteq \{0, 1, \ldots, m-1\}$, and $0 \in S$.  We
will now describe a polynomial-time algorithm that decides $\uc{S}{m}$.  Let $G$
be the input graph.  Check, using algorithm in Theorem~\ref{thm:tree-width}, if
$G$ has tree-width at least $t_m$, where $t_m$ is as in
Theorem~\ref{thm:large-tree-width}.  If so, then, by
Theorem~\ref{thm:large-tree-width}, $G$ has a cycle of length $\mymod{0}{m}$.
Otherwise, $G$ has tree-width at most $t_m$.  So, we use
Theorem~\ref{thm:k-disj-paths} to check if $G$ has a cycle of length $\ell$ such
that $\ell \in S$.  For all distinct vertices $v_1, v_2, v_3, v_4$ in $G$ such
that $\{(v_1, v_2), (v_3,v_4)\} \subseteq E(G)$, for each $\ell \in S$, and for
each $0 \leq d_1, d_2 \leq m-1$ such that $d_1 + d_2 + 2 = \mymod{\ell}{m}$, we
do the following.  Check, using Theorem~\ref{thm:k-disj-paths}, whether there
are 2 disjoint paths $P_1$ and $P_2$, $P_1$ between $v_1$ and $v_3$ of length
$\mymod{d_1}{m}$ and $P_2$ between $v_2$ and $v_4$ of length $\mymod{d_2}{m}$.
If there are such disjoint paths, then there is a cycle of length $d_1 + d_2 + 2
= \mymod{\ell}{m}$, namely the cycle consisting of the edges in $P_1$, the edges
in $P_2$, and the edges $(v_1, v_2)$ and $(v_3, v_4)$.  Note that we may be
missing cycles consisting of 3 nodes or less, but that can be easily handled by
checking brute-force for all cycles of 3 nodes or less.
%\qed

\medskip
%%% several ways to prove ({1..m-1},m)-UC is in P
%%% 1. note that ({1,2,...m-1}, m)-UC can be done easily using APY's result that finding
%%% the period of an undirected (as also a directed) graph is in P
%%% 2. corollary \ref{cor:uc-all-except-0} 
%%% 3. modifying yuster and zwick will result in actually finding the shortest such path

Let us consider the complements of the following cycle length modularity problems (for fixed $m \geq
2$ and fixed $0 \leq r < m$): $\uc{\{0,1,\ldots,m-1\}-\{r\}}{m}$.  For any $m$
and $r$ such that $0 \leq r < m$, these problems asks whether all cycles in the
given graph are of length $\equiv \mymod{r}{m}$.  For $m = 2$ and $r = 0$, this
problem is the odd cycle problem in undirected graphs, which as noted in the
introduction is easily seen to be in $\p$ based on the simple observation that
any closed walk of odd length in a graph must contain a simple cycle of odd
length.  For $m = 2$ and $r = 1$, this problem is the even cycle problem, which
is also in $\p$~\cite{ark-pap-yan:j:modularity}.  Arkin, Papadimitriou, and
Yannakakis in fact prove, via using the properties of triconnected components of
graphs, that for each $m$, and each $0 \leq r < m$, finding whether all cycles
in a graph are of length $\equiv \mymod{r}{m}$ can be done in polynomial time.
\begin{theorem}[\cite{ark-pap-yan:j:modularity}]
\label{thm:uc-all-except-one}
For each $m \in \naturalnumber$, and each $r$ such that $0 \leq r < m$,
$\uc{\{0,1,\ldots,m-1\}-\{r\}}{m} \in \p$.
\end{theorem}
\begin{corollary}\label{cor:uc-all-except-0}
  $\uc{\{1,2,\ldots,m-1\}}{m}\in \p$.
\end{corollary} 

\noindent
The following theorem is an analog of Theorem~\ref{thm:dir-0-out}(ii) for
undirected graphs.
\begin{theorem} \label{thm:undir-0-out}
  For all $m>2$ and $S\subseteq \{1, \ldots , m-1\}$, the
  following is true: If for all $p\in S$, and all $d_1, d_2$, such that $0\le d_1,
  d_2 < m$ and $d_1 + d_2\equiv \mymod{p}{m}$, it holds that $d_1\in S$ or $d_2\in
  S$, then $\uc{S}{m}\in\p$.
\end{theorem}
The proof given for the corresponding statement regarding directed graphs does not
work here. The reason is that closed walks in undirected graphs need not decompose properly
into cycles.  To see why this is true, consider a closed walk $C$ of length $5$ in
an undirected graph: $v_1v_2v_3v_4v_2v_1$.  Note that even though $C$ is a
closed walk of length $5$, it is neither a cycle nor does it decompose properly into
cycles, basically because $v_1v_2v_1$ is not a valid cycle. 

In order to prove Theorem~\ref{thm:undir-0-out}, we reduce the problem to the problem of determining the period of
an undirected graphs, which is solvable in polynomial time by the algorithm from
Arkin, Papadimitriou, and Yannakakis~\cite{ark-pap-yan:j:modularity}.
We need the following lemma.
\begin{lemma} \label{lem::auxlemma}
  For all $m\ge 1$ and $S = \{ a_1, \ldots, a_n\} \subseteq \{0,\ldots , m-1\}$, $S \neq \emptyset$,
  the following is true:\\
  If for all $d_1\in S, d_2\in S$ it holds that 
  $(d_1 + d_2) \bmod m \in S$, then $S = \{ \ell \condition 0\le \ell < m \mbox{ and }
  g|\ell\}$ for some $g$ with $g|m$.
\end{lemma}
\sproof
% First note that $0 \in S$: let $a_1 \in S$, Then, $(ma_1) mod m = 0 \in S$.
Let $S - \{0\} = \{a_1,a_2,\ldots, a_n\}$
Let $g = \gcd(a_1, \ldots , a_n, m)$. From number theory (see~\cite{apo:b:number})
we know that there exist $k_1, \ldots , k_n, k_{n+1} \in \integers$, such that
\begin{displaymath}
   k_1a_1 + k_2a_2 + \cdots + k_na_n + k_{n+1}m = g.
\end{displaymath}
For all $i$, $1\le i \le n+1$, let
\begin{displaymath}
  k_i' \stackrel{df}{=} k_i + m|k_i|.
\end{displaymath}
Then
\begin{equation} \label{eq::ggt}
  (k_1'a_1 + \cdots + k_n'a_n) \bmod m = g,
\end{equation}
where $k_1', \ldots , k_n' \ge 0$.

\noindent
For all $d_1\in S$, $d_2\in S$ it holds that $(d_1 + d_2) \bmod m \in S$.
Hence Eq.~(\ref{eq::ggt}) implies that $g\in S$. Furthermore, $sg\in S$ for all $s\in\integers$ such that 
$0\le sg < m$.

\noindent
Since for each $1 \leq i \leq n$, $g|a_i$, it follows that
\begin{displaymath}
  S = \{ \ell \condition 0\le \ell < m \mbox{ and } g|\ell \}.
\end{displaymath}
%\eproofof{Lemma~\ref{lem::auxlemma}}
\noindent This concludes the proof of Lemma~{\ref{lem::auxlemma}

\medskip

\noindent {\bf Proof of Theorem~\ref{thm:undir-0-out}.}
Let $\overline{S} \stackrel{df}{=} \{ 0, \ldots , m-1\} - S$. 
Lemma~\ref{lem::auxlemma} implies that
\begin{displaymath}
  \overline{S} = \{ \ell \condition 0\le \ell < m \mbox{ and } g| \ell \}
\end{displaymath}
for some $g$ with $g|m$.
Hence
\begin{displaymath}
   S = \{ \ell \condition 0\le \ell < m \mbox{ and } g\not| \ell \}.
\end{displaymath}
Define 
\begin{displaymath}
   S'= \{ 1, \ldots , g-1\}. 
\end{displaymath}
Since $g|m$ holds
\begin{displaymath}
  x\bmod m \in S \Longleftrightarrow x\bmod g \in S' 
\end{displaymath}
for all $x\in\naturalnumber$.
Hence $\uc{S}{m}$ is equivalent to $\uc{\{1,\ldots , g-1\}}{g}$. 
However, $\uc{\{1,\ldots , g-1\}}{g}$ is the set of graphs containing a cycle not divisible by $g$,
which is in $\p$ since the period of a graph (the $\gcd$ of all cycle lengths) can be determined
in polynomial time~\cite{ark-pap-yan:j:modularity}.  This concludes the proof of Theorem~\ref{thm:undir-0-out}.

\section{Conclusion and Open Problems}\label{sec:conclusion}
In this paper, we studied the complexity of cycle length modularity problems.
We completely characterized (i.e., as either polynomial-time computable or as
$\np$-complete) each problem $\dc{S}{m}$, where $0 \notin S$.  We also proved
that, for each $S$ such that $0 \in S$, $\uc{S}{m}$ is in $\p$, and we proved a
sufficient condition on $S$ and $m$ for the problem $\uc{S}{m}$ ($0 \notin S$)
to be in $\p$.
We mention
several open problems.
\begin{enumerate}
\item Theorem~\ref{thm:dir-0-out} completely characterizes all modularity
problems in directed graphs when $0 \notin S$.  Robertson, Seymour, and
Thomas~\cite{rob-sey-tho:j:pfaffian} prove that $\dc{\{0\}}{2}$ is in $\p$.  In
light of these results, it is natural to ask if $\dc{\{0\}}{m} \in \p$, for some
or all $m > 2$.  Also, the complexity of $\dc{S}{m}$ such that $0 \in S$ and $m
> 2$ is still open, except for trivial ($S = \{0,1,\ldots,m-1\}$) cases.
\item Theorem~\ref{thm:ucsm-0-in} shows that all cycle length modularity
problems in undirected graphs $\uc{S}{m}$ such that $0 \in S$ are solvable in
polynomial time.  
What about the complexity of the $\uc{S}{m}$ problems with $0 \notin S$
which are not covered by Theorem 
\ref{thm:uc-all-except-one} or \ref{thm:undir-0-out}?

%What about the complexity of $\uc{S}{m}$ where $0 \notin S$?
%Note that if $S$ is such that $||S|| = m-1$ then,
%by~\cite{ark-pap-yan:j:modularity}, we get that $\uc{S}{m}$ is in $\p$, but all
%the remaining problems are still open.

\item Theorem~\ref{thm:uc-all-except-one} shows that, for undirected graphs, the
problem of finding whether all cycles have length $\equiv \mymod{r}{m}$ is in
$\p$.  %However, we do not have a corresponding result for the case of directed
%graphs.  In Corollary~\ref{cor:dc-all-except-0}, we have a partial result,
%namely that, for directed graphs, the problem of deciding whether all cycles
%have length $\equiv \mymod{0}{m}$ is in $\p$.  But the general problem of
%$\dc{\{0,1,\ldots,m-1\}-\{r\}}{m}$ is still open.
What is the complexity of the corresponding problem for directed graphs?
\end{enumerate}

\bibliography{grythakur}

\begin{thebibliography}{MRST97}

\bibitem[Apo76]{apo:b:number}
T.~Apostol.
\newblock {\em Introduction to Analytic Number Theory}.
\newblock Undergraduate Texts in Mathematics. Springer-Verlag, 1976.

\bibitem[APY91]{ark-pap-yan:j:modularity}
E.~Arkin, C.~Papadimitriou, and M.~Yannakakis.
\newblock Modularity of cycles and paths in graphs.
\newblock {\em Journal of the ACM}, 38(2):255--274, April 1991.

\bibitem[BV73]{bal-vei:j:period}
Y.~Balcer and A.~Veinott.
\newblock Computing a graph's period quadratically by node condensation.
\newblock {\em Discrete Mathematics}, 4:295--303, 1973.

\bibitem[EP65]{erd-pos:j:independent-circuits}
P.~Erd{\"{o}}s and L.~P{\'{o}}sa.
\newblock On independent circuits contained in a graph.
\newblock {\em Canadian Journal on Mathematics}, 17:347--352, 1965.

\bibitem[FHW80]{for-hop-wyl:j:subgraph-homeomorphism}
S.~Fortune, J.~Hopcroft, and J.~Wyllie.
\newblock The directed subgraph homeomorphism problem.
\newblock {\em Theoretical Computer Science}, 10:111--121, 1980.

\bibitem[GL96]{gal-loe:j:prescribed}
A.~Galluccio and M.~Loebl.
\newblock Cycles of prescribed modularity in planar digraphs.
\newblock {\em Journal of Algorithms}, 21:51--70, 1996.

\bibitem[Kas67]{kas:b:graph-theory-crystal-physics}
P.~Kasteleyn.
\newblock Graph theory and crystal physics.
\newblock In F.~Harary, editor, {\em Graph Theory and Theoretical Physics},
  pages 43--110. Academic Press, New York, 1967.

\bibitem[Knu73]{knu:tr:strong-components}
D.~Knuth.
\newblock Strong components.
\newblock Technical Report 004639, Computer Science Department, Stanford
  University, Stanford, California, 1973.

\bibitem[MRST97]{mcc-rob-sey-tho:c:pfaffian}
W.~McCuaig, N.~Robertson, P.~Seymour, and R.~Thomas.
\newblock Permanents, pfaffian orientations, and even directed circuits.
\newblock In {\em Proceedings of the 29th ACM Symposium on Theory of
  Computing}, pages 402--405, 1997.

\bibitem[Pol13]{pol:j:aufgabe}
G.~Polya.
\newblock Aufgabe 424.
\newblock {\em Arch. Math. Phys.}, 20(3):271, 1913.

\bibitem[RS85]{rob-sey:b:survey}
N.~Robertson and P.~Seymour.
\newblock Graph minors---a survey.
\newblock In I.~Anderson, editor, {\em Surveys in Combinatorics 1985: Invited
  Papers for the Tenth British Combinatorial Conference}, pages 153--171.
  Cambridge University Press, 1985.

\bibitem[RS86]{rob-sey:j:graph-minors-2}
N.~Robertson and P.~Seymour.
\newblock Graph minors {I}{I}. {A}lgorithmic aspects of tree-width.
\newblock {\em Journal of Algorithms}, 7:309--322, 1986.

\bibitem[RST99]{rob-sey-tho:j:pfaffian}
N.~Robertson, P.~Seymour, and R.~Thomas.
\newblock Permanents, pfaffian orientations, and even directed circuits.
\newblock {\em Annals of Mathematics}, 150:929--975, 1999.

\bibitem[Tar72]{tar:j:dfs}
R.~Tarjan.
\newblock Depth first search and linear graph algorithms.
\newblock {\em SIAM Journal on Computing}, 2:146--160, 1972.

\bibitem[Tho88]{tho:j:disjoint-subgraphs}
C.~Thomassen.
\newblock On the presence of disjoint subgraphs of a specified type.
\newblock {\em Journal of Graph Theory}, 12(1):101--111, 1988.

\bibitem[VY89]{vaz-yan:j:pfaffian}
V.~Vazirani and M.~Yannakakis.
\newblock Pfaffian orientations, 0-1 permanents, and even cycles in directed
  graphs.
\newblock {\em Discrete Applied Mathematics}, 25:179--190, 1989.

\bibitem[YZ97]{yus-zwi:j:even-cycles}
R.~Yuster and U.~Zwick.
\newblock Finding even cycles even faster.
\newblock {\em SIAM Journal on Discrete Mathematics}, 10(2):209--222, May 1997.

\end{thebibliography}

\end{document}